# Reemergeing electronic nematicity in heavily hole-doped Fe-based superconductors

J. Li[1], D. Zhao[1], Y. P. Wu[1], S. J. Li[1], D. W. Song[1], L. X. Zheng[1], N. Z. Wang[1], X. G. Luo[1,2,4], Z. Sun[2,3,4], T. Wu[1,2,4] and X. H. Chen[1,2,4]

1. Hefei National Laboratory for Physical Sciences at Microscale and Department of Physics, University of Science and Technology of China, Hefei, Anhui 230026, China

2. Key Laboratory of Strongly-coupled Quantum Matter Physics, Chinese Academy of Sciences, School of Physical Sciences, University of Science and Technology of China, Hefei, Anhui 230026, China

3. National Synchrotron Radiation Laboratory, University of Science and Technology of China, Hefei, Anhui 230026, China

4. Collaborative Innovation Center of Advanced Microstructures, Nanjing University, Nanjing 210093, China

**In correlated electrons system, quantum melting of electronic crystalline phase often gives rise to many novel electronic phases. In cuprates superconductors, melting the Mott insulating phase with carrier doping leads to a quantum version of liquid crystal phase, the electronic nematicity, which breaks the rotational symmetry and exhibits a tight twist with high-temperature superconductivity[1]. Recently, the electronic nematicity has also been observed in Fe-based superconductors[2]. However, whether it shares a similar scenario with its cuprates counterpart is still elusive[3]. Here, by measuring nuclear magnetic resonance in $CsFe_2As_2$, a prototypical Fe-based superconductor perceived to have evolved from a Mott insulating phase at $3d^5$ configuration[4], we report anisotropic quadruple broadening effect as a direct result of local rotational symmetry breaking. For the first time, clear connection between the Mott insulating phase and the electronic nematicity can be established and generalized**

**to the Fe-based superconductors. This finding would promote a universal understanding on electronic nematicity and its relation with high-temperature superconductivity.**

Electronic nematicity, a quantum version of liquid crystal phase, spontaneously breaks rotational symmetry of the underlying crystal lattice. In Fe-based superconductors (FeSCs), the existence of nematic order is a well-established experimental fact[3]. As more carriers are doped, a nematic quantum critical point (QCP) is revealed near the optimal doping as shown in Fig. 1. Hence, nematic fluctuation is believed to be important for high-temperature superconductivity in FeSCs[2,3,5]. However, the origin and role of such nematicity remains controversial[3]. In cuprates superconductors, electronic nematicity is a phenomenological consequence of quantum melting of Mott insulating phase by carrier doping[1], which has been suggested as a possible source of the pseudogap phase[6-8]. Whether the electronic nematicity in FeSCs would also arise from a similar scenario is a fundamentally important issue. Up to date, the connection between electronic nematicity and Mott insulating phase is still elusive. Very recently, a heavy-fermion-like behavior was observed in heavily hole-doped FeSCs $AFe_2As_2$ (A = K, Rb, Cs)[4,9-11], indicating a possible orbital-selective Mott transition. Moreover, as shown in Fig.1, the electron correlation deduced from effective mass shows a monotonic increase from a Fermi-liquid phase at heavy electron doping[4]. Assuming a half-filling Mott insulating phase at $3d^5$ configuration, the heavily hole-doped FeSCs $AFe_2As_2$ (A = K, Rb, Cs) at $3d^{5.5}$ configuration could be treated as a doped Mott insulator with 10% doping per each orbital[4]. Whether the electronic nematicity exists in this regime will be a direct test for the potential link between electronic nematicity and the Mott insulating phase. On the other hand, electronic nematicity has never been observed in heavily hole or electron doping regime before[12]. Here, by measuring nuclear magnetic resonance (NMR), we show that the electronic nematicity indeed exists, albeit in a different form, in such regime for the first time.

As shown in Fig. 2a and 2e, we measured the NMR spectrum of $^{75}As$ nuclei with the magnetic field parallel and perpendicular to the c axis, respectively. $^{75}$As (nuclear spin I = 3/2) NMR spectrum is split into one central peak and two satellite peaks due to the nuclear quadrupole interaction with electric field gradient (EFG) produced by environment. Under the first-order approximation, the resonance frequency of the central peak is $f_\alpha^c = (1 + K_\alpha)\gamma_n B_\alpha$ ($\alpha = a, b, c$) and that of satellite peak is $f_\alpha^s = (1 + K_\alpha)\gamma_n B_\alpha \pm V_\alpha$ ($\alpha = a, b, c$). $\gamma_n$ = 7.2919 MHz/T is the nuclear gyromagnetic ratio. $K_\alpha$ is the Knight shift, which measures the product between the local spin susceptibility and the hyperfine coupling constant. $V_\alpha$ is the principal component of the EFG tensor. In the local crystal geometry as shown in Fig. 2a, four Fe atoms and one $^{75}$As atom forms a pyramid structure, with the $^{75}$As atom sitting on the apex. Such configuration grants extreme sensitivity to the apical $^{75}$As to any in-plane rotational symmetry breaking effect.

Usually, when in-plane rotational symmetry is broken with a spontaneous nematic domain texture, the NMR spectrum under in-plane magnetic field would exhibit a small splitting on central peak, such as in the case of FeSe[13,14] and LaOFeAs[15]. Such splitting has two kinds of physical origin. One is from the anisotropy of Knight shift ($K_a \neq K_b$) and the other is from the anisotropy of second order effect of nuclear quadrupole interaction with EFG. Both of anisotropies are the manifestation of electronic nematicity[13-15]. However, on the satellite peaks, above splitting effect is negligible and the dominating splitting effect shall directly originate from the anisotropy of in-plane EFG tensor ($V_a \neq V_b$) due to rotational symmetry breaking, such as in the case of P-doped BaFe$_2$As$_2$[16] and Co-doped NaFeAs[17]. Therefore, it is the satellite peak that has great advantage to detect the manifestation of electronic nematicity from orbital or charge degree of freedom.

Here, no distinguishable splitting is observed on both central peak and satellite peaks throughout whole temperature range down to 2 K, suggesting the absence of nematic ordering. This is consistent with the absence of any magnetic and structural transition in this system[18]. However, once a magnetic field is applied in-plane, remarkable spectral broadening was observed on the satellite peaks as indicated in Fig.

2b and 2d. The EFG effect is shown to dominate in this situation given the significantly larger broadening on the satellite peaks than that on the central peak. In comparison, an out-of-plane magnetic field yields weaker broadening effect, while the EFG effect still dominates. The temperature-dependent linewidths for both central peak and satellite peak are summarized in Fig. 2i and 2j.

The anisotropic broadening effect can be understood quantitatively in term of EFG tensor $V_\alpha$. In general, the principal components of the EFG tensor must satisfy the Laplace relationship of $V_a + V_b + V_c = 0$. Therefore, the EFG broadening should also follow $\Delta V_a + \Delta V_b + \Delta V_c = 0$. When in-plane rotational symmetry remains with $V_a = V_b$, the anisotropic EFG broadening follows $\Delta V_a/V_a = \Delta V_c/V_c$. When in-plane rotational symmetry breaks with $V_a \neq V_b$, a spontaneous nematic domain texture pinned by disorder would lead to an additional contribution to $\Delta V_a/V_a$, which breaks $\Delta V_a/V_a = \Delta V_c/V_c$. This effect has already been observed in previous NMR study on disorder pinned nematicity[16,17]. In this case, $\Delta V_a/V_a = \Delta V_c/V_c + |V_a - V_b|/V_a$ and the additional contribution from $|V_a - V_b|/V_a$ is related to the mean value of in-plane asymmetric parameter defined by $\eta = |(V_a - V_b)/V_c| = (\Delta V_a/V_a - \Delta V_c/V_c)/2$ which is a good indicator of electronic nematicity. As shown in Fig. 2k, the observed $\Delta V_a/V_a$ and $\Delta V_c/V_c$ clearly breaks the required relationship by the conservation of in-plane rotational symmetry and supports a definitive contribution from in-plane asymmetric parameter $\eta$, which unambiguously indicates a locally broken in-plane rotational symmetry in the Fe-As plane.

Moreover, the extracted in-plane asymmetric parameter $\eta$ shows a strong temperature-dependence. The temperature-dependent $\eta$ is plotted in Fig. 3. Above 20 K, the temperature-dependent $\eta$ follows a standard Curie-Weiss behavior with Weiss temperature $\theta \sim -20$ K. Similar Curie-Weiss behavior has been widely observed for temperature-dependent nematic susceptibility ($\chi_{nem}$) of FeSCs above the nematic ordering temperature[2,5,19]. As the Weiss temperature crosses to zero, a nematic QCP would emerge at low temperature[5,19]. Given the linear dependence between $\eta$ and $\chi_{nem}$, both quantities should follow the same Curie-Weiss behavior (See supplementary information). The negative Weiss temperature θ as fitted in this case is consistent with the lack of explicit nematic ordering down to 2K, which has also been evidenced by other FeSCs[2,5,19].

Below 20 K, the temperature-dependent η deviates from the high-temperature Curie-Weiss behavior with a distinct plateau behavior. Similar deviation from Curie-Weiss behavior has been also observed in previous study on nematic QCP[5,19]. In that case, the quantum criticality would be affected by disorder and the deviation from Curie-Weiss behavior could be a manifestation of disorder effect on quantum criticality[5,19]. Here, the absence of magnetic and structural transition with a small negative Weiss temperature strongly suggests a similar nematic QCP behavior. In fact, disorder effect already shows up in high-temperature Curie-Weiss region with a local pinning effect of nematic fluctuation and triggers a short-range nematic ordering phenomenon to broaden the lineshape of NMR spectrum, which shares a great similarity with disorder pinned charge fluctuation in underdoped YBCO[20].

On the other hand, the difference of the spin dynamics between the plateau region and the Curie-Weiss region is also investigated. As shown in the insets of Fig. 3, we measured the spin-lattice relaxation ($T_1$) of $^{75}$As nuclei in these two regions. Above 20 K, the spin-lattice relaxation measurement shows a standard spin-echo decay process with single $T_1$ component. In contrast, below 20 K, a distinct spin-echo decay process with two $T_1$ components appears, which strongly supports an exotic spin dynamics in plateau regime.

In Fig. 4, both temperature and field dependence of above two-component spin-lattice relaxation are measured. It is figured out that the two-component spin-lattice relaxation below 20 K is actually a field-induced phenomenon. When we perform spin-lattice relaxation measurement with zero external field (nuclear quadrupole resonance, NQR), the spin-echo decay process with single $T_1$ component is recovered as shown in Fig. 4b. As shown in Fig. 4d, with increasing magnetic field up to 12 Tesla, the second component of spin-lattice relaxation gradually comes out and coexists with the zero-field component. The ratio of these two $T_1$ components at 12 Tesla is close to 1:1. As shown in Fig. 4a, the two $T_1$ components under 12 Tesla show different temperature dependence from that of NQR measurement[10] with $1/T_1^{NQR} \sim T^{0.75}$. The longer component $T_{1L}$ follows a Fermi-liquid-like behavior with $1/T_{1L} \sim T$ and the shorter component $T_{1s}$ follows a non-Fermi-liquid-like behavior with $1/T_{1s} \sim T^{0.35}$. A similar field-induced two-component spin-lattice relaxation was recently observed near the QCP, such as in Kondo-lattice compound $YbRh_2Si_2$[21]. In comparison, the magnetic susceptibility in $YbRh_2Si_2$ is the analog of the nematic susceptibility in $CsFe_2As_2$ represented by η. Interestingly, in $YbRh_2Si_2$, the magnetic susceptibility also shows a similar plateau behavior when two-component feature appears[21]. Considering the similarity between these two systems, the field-induced two-component spin-lattice relaxation in $CsFe_2As_2$ might be also ascribed to an intrinsic QCP feature as that in $YbRh_2Si_2$. A similar two-component behavior of spin-lattice relaxation was also reported in $RbFe_2As_2$ polycrystalline samples[22].

Our observation offers an unprecedented opportunity to unravel the origin of the nematicity. As shown in Fig. 1, the orientation of nematic domain observed in present work is revealed to be along Fe-As-Fe direction (See supplementary information) instead of Fe-Fe direction in previous studies[13-17]. There is a π/4 phase change between these two kinds of electronic nematicity, which makes them conflict with each other. This also shed light on the disappearance of electronic nematicity at moderate hole-doped $BaFe_2As_2$[12], which could be ascribed to the competition between above two conflicting electronic nematicities. On the other hand, the evolution of electronic nematicity seems to be qualitatively consistent with the evolution of the low-energy spin excitation revealed by previous inelastic neutron scattering[23]. The detail of the low-energy spin excitation is important to determine the orientation of nematic domain. When the dispersive low-energy spin excitation is wiped out in heavily electron-doped regime, the electron nematicity fades out and a Fermi liquid state emerges. This suggests a key role of magnetic fluctuation on nematicity.

In addition, the present work also builds up a perfect connection between electronic nematicity and Mott insulating phase. Here, due to the limitation of materials, the Mott insulating phase is still a theoretical expectation[4]. As shown in Fig. 1, although the exact phase boundary of electronic nematicity at $3d^{5.5}$ configuration is still unclear at present stage, our finding solves a pivotal missing piece in the global phase diagram which exhibits a continuous quantum melting of Mott insulator at half-filling $3d^5$ configuration into a Fermi liquid state. Under this scenario, the electronic nematicity in FeSCs is also a phenomenological consequence of the melting of Mott insulator as cuprates. This would fundamentally unify the origin of electronic nematicity and also pairing mechanism in both high-Tc families. In cuprates, significant charge fluctuation or charge order has been widely observed in pseudogap region[24-28]. In comparison, charge fluctuation might be also important for electronic nematicity in FeSCs. In fact, a simple simulation based on point charge model easily yields a consistent value for in-plane asymmetric parameter η shown in Fig. 3 with ~$10^{-3}$ e intra-unit-cell charge imbalance (See supplementary information). Very recently, two possible signatures of charge order have also been reported in $RbFe_2As_2$[22] and pressurized $KFe_2As_2$[29] by NQR and NMR measurement. Whether charge order is intrinsic property or not still needs further investigation.

Finally, electronic nematicity has been an ubiquitous physical property in high-temperature superconductors, which exhibits a complicated entanglement with other novel forms of quantum matter in phase diagram, such as high-temperature superconductivity, pseudogap phase, stripe order, charge order and so on[30]. How to understand its underlying physical origin would be a key to understand the complexity of the phase diagram of high-temperature superconductors. Our finding would promote a universal understanding on electronic nematicity, and also shed light on the underlying physics of high-temperature superconductivity.

**Methods**

Methods are available in supplementary information.

**Acknowledgments**

We thank G. M. Zhang, Y. He, M. -H. Julien and J. Schmalian for valuable discussions. This work is supported by the National Key R&D Program of the MOST of China (Grant No. 2016YFA0300201), the National Natural Science Foundation of China (Grants No. 11522434, 11374281, U1532145), the "Strategic Priority Research Program (B)" of the Chinese Academy of Sciences (Grant No. XDB04040100), the Fundamental Research Funds for the Central Universities and the Chinese Academy of Sciences. T. W. acknowledges the Recruitment Program of Global Experts and the CAS Hundred Talent Program.

**Additional Information**

The authors declare no competing financial interests. Correspondence and requests for materials should be addressed to T. W. (wutao@ustc.edu.cn).

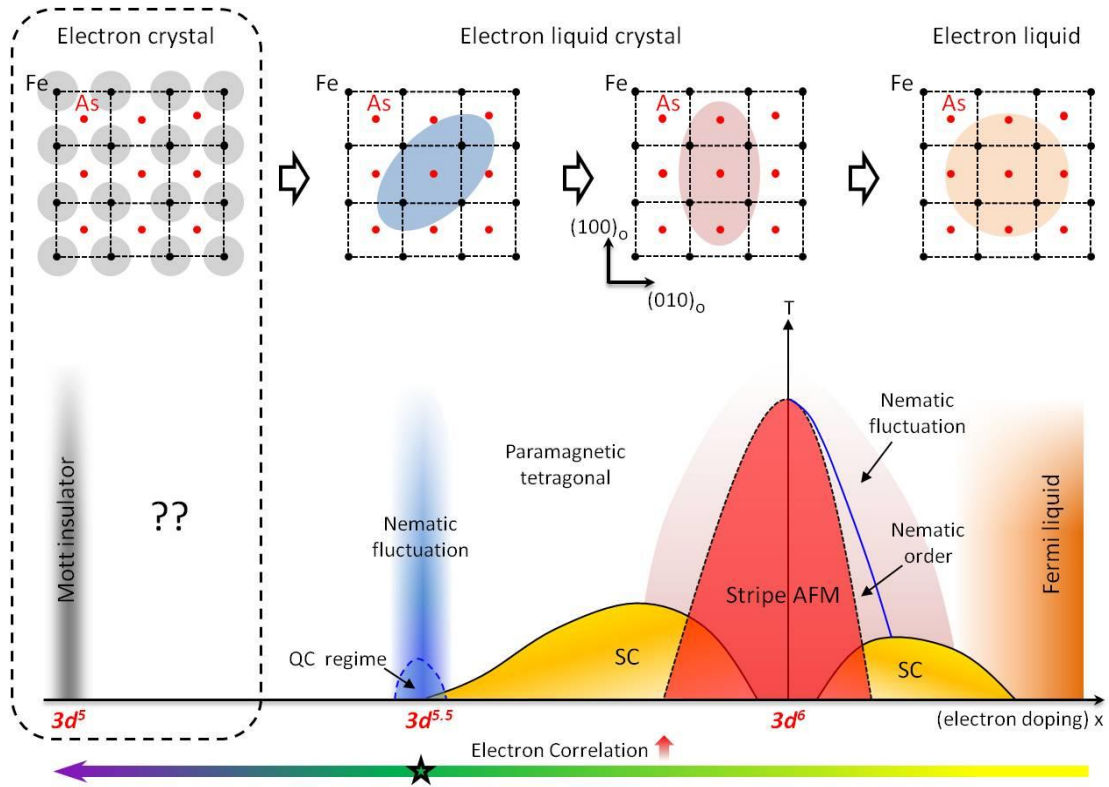

**Figure 1 | Global phase diagram of FeSCs with 122-type structure.** The red area denotes stripe-type antiferromagnetism, the area below solid blue line denotes nematic/orthorhombic paramagnetic order, and the yellow area denotes superconductivity. The reddish area denotes a fluctuating nematic phase around $3d^6$ configuration with nematic domain along Fe-Fe direction. The area below the blue dash line denotes a possible quantum criticality regime affected by disorder. The blueish area denotes a fluctuating nematic phase at $3d^{5.5}$ configuration with nematic domain along Fe-As-Fe direction. The orange area denotes a Fermi liquid phase. The gray area denotes a Mott insulator at $3d^5$ configuration. The arrow denotes the increasing of electron correlation deduced from effective mass[4]. The area encircled by black dash line is determined by theoretical prediction. The other area is determined by experiments. The pentagram represents the region studied in the present work. The upper insets depict the different quantum states of matter in real space which corresponds to Mott insulator (in gray area), electron liquid crystal (in reddish and blueish areas) and electron liquid (in orange area).

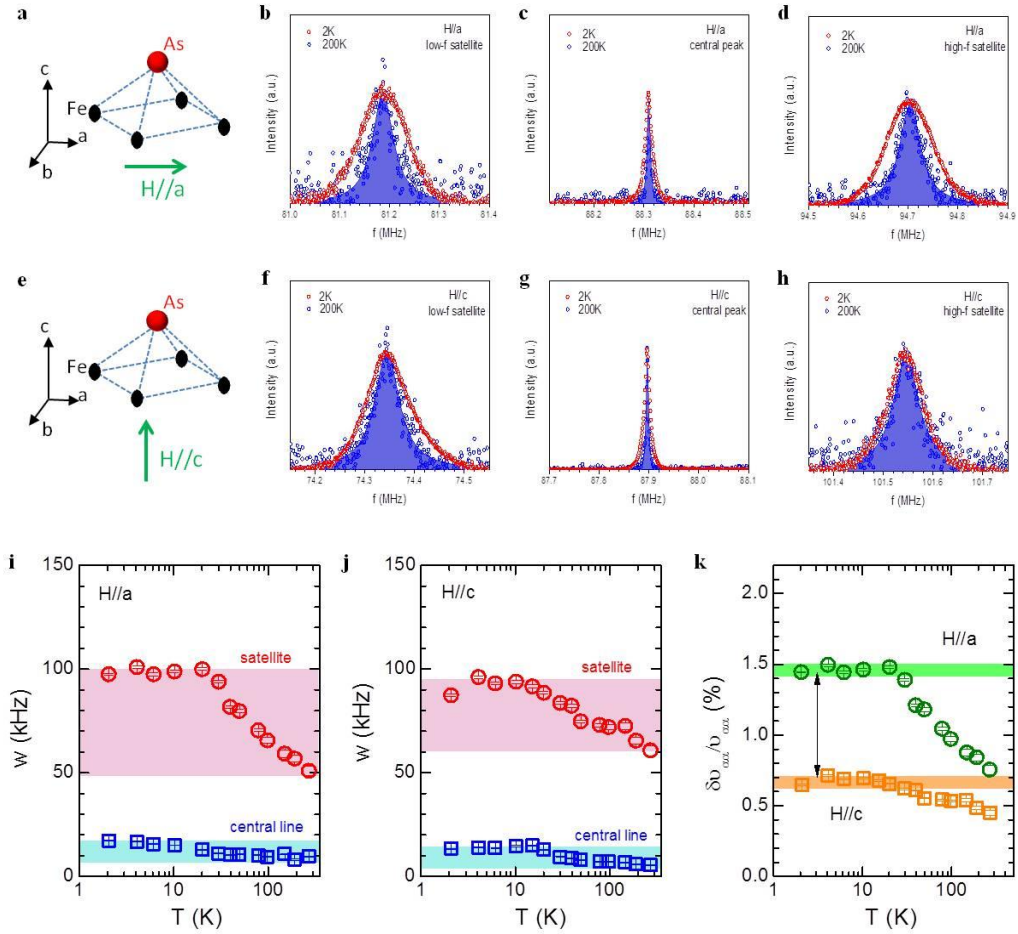

**Figure 2 | NMR Evidence for in-plane rotational symmetry breaking in $CsFe_2As_2$.** (**a**) and (**e**) are the sketches of two NMR experimental configurations. (**b**)-(**d**) and (**f**)-(**h**) are the measured spectra of the $^{75}As$ nuclei with the external magnetic field (H = 12 T) applied parallel to a-axis and c-axis, respectively. The red and blue open circles denote the original experimental data at 2 K and 200 K, respectively. The blue areas are the fitting results for the experimental data at 200 K. The spectra measured at 2K and 200K are normalized and overlapped together to make a contrast. The low-f satellite, high-f satellite and central peak are plotted separately with a same frequency scaling. There is a tiny asymmetric broadening effect at 2 K when H // c. It could be ascribed to a twisted broadening effect with both magnetic and quadrupole origins due to charge modulation[20,26]. Here, this is a secondary phenomenon and we leave it for future work. (**i**) and (**j**) show the temperature dependence of linewidth of the central peak and averaged linewidth of the low- and high- frequency satellites. It is obviously that the averaged linewidth of satellites under H // a exhibits a stronger broadening effect with temperature cooling in compared with the situation under H//c. (**k**) the comparison of the normalized linewidth of the satellite with H // a and H // c. It apparently breaks the relationship $\Delta V_a/V_a = \Delta V_c/V_c$ when the in-plane rotational symmetry remains, which could be owed to an extra linewidth broadening effect caused by the in-plane rotational symmetry breaking as explained in the text.

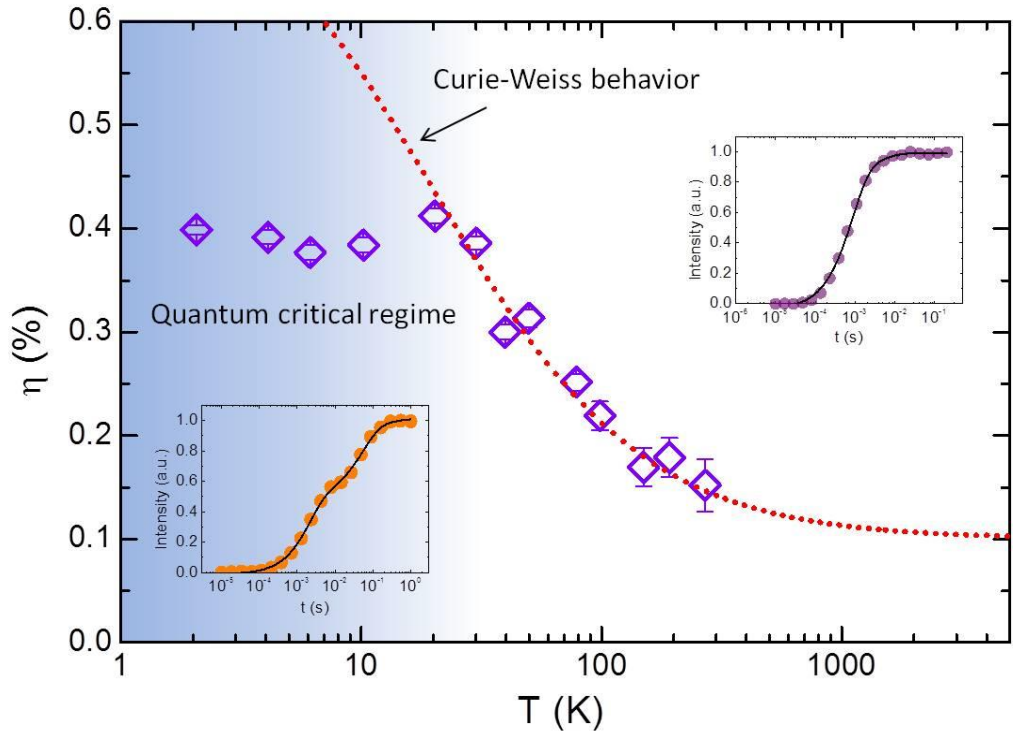

**Figure 3 | Temperature-dependent in-plane asymmetric parameter of EFG.** The data marked as purple rhombus are reduced from the data in Fig. 2K. The details are described in the main text. The red dash line is a Curie-Weiss fitting of the η ~ T relation. The Curie-Weiss fitting gives a reasonable description of the observed results above 20 K. Below 20 K, the system is believed to enter a disorder affected quantum critical regime (denoted by blueish area) and η shows a distinct plateau behavior. The insert at the top right corner shows the single-component $T_1$ decay process of the NMR spin-lattice relaxation at 50 K. While the insert at the left bottom corner shows the two-component $T_1$ behavior at 2 K when the system enter the disorder affected quantum critical regime. The spin-lattice relaxation measurement by NMR is conducted on central peak.

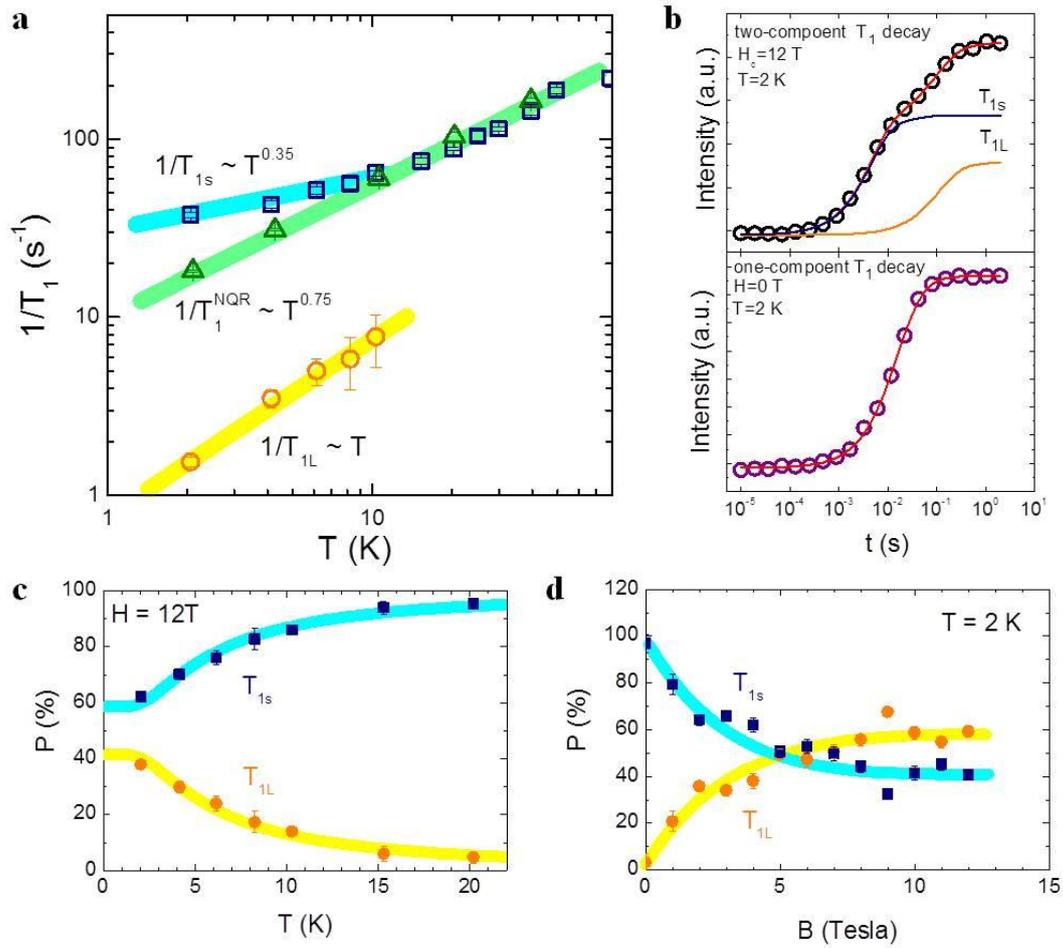

**Figure 4 | Exotic spin dynamics in quantum critical regime.** (**a**) Temperature dependence of the two-component $1/T_1$ (shorter component $1/T_{1S}$, longer component $1/T_{1L}$) under out-of-plane magnetic field $H_c = 12$ T and the single-component $1/T_1^{NQR}$ measured under zero-field. The two-component behavior in spin-lattice relaxation gradually disappears above 20 K and the longer component $1/T_{1S}$ becomes negligible above 10 K. Therefore, only shorter component $1/T_{1S}$ is plotted above 10 K. The bold color lines are guiding for the eyes. The spin-lattice relaxation measurement by NMR is conducted on central peak. (**b**) Comparison of the low temperature (T = 2 K) two-component decay process under out-of-plane magnetic field $H_c = 12$ T and single-component decay process under zero field. In the top panel of figure (**b**) the separated decay process of shorter $T_{1s}$ component and longer $T_{1L}$ component are depicted as blue and orange solid lines. The red lines are the fitting results. (**c**) Temperature dependence of the intensity weight of the two $T_1$ components. (**d**) External magnetic field dependence of the intensity weight of the two $T_1$ components. The bold color lines are guiding for the eyes. The field-dependent spin-lattice relaxation measurement by NMR is conducted on high-f satellite peak.